\documentclass[prl,twocolumn,a4paper,showpacs,superscriptaddress]{revtex4-1}
\usepackage{amsmath}
\usepackage{amsfonts}
\usepackage{graphicx}
\begin{document}
\title{Macrorealism from entropic Leggett-Garg inequalities}
\author{A. R. Usha Devi}
\email{arutth@rediffmail.com}
\affiliation{Department of Physics, Bangalore University, 
Bangalore-560 056, India}
\affiliation{Inspire Institute Inc., Alexandria, Virginia, 22303, USA.}
\author{H. S. Karthik}
\affiliation{Raman Research Institute, Bangalore 560 080, India}
\author{Sudha} 
\affiliation{Department of Physics, Kuvempu University, Shankaraghatta, Shimoga-577 451, India.}
\affiliation{Inspire Institute Inc., Alexandria, Virginia, 22303, USA.}
\author{A. K. Rajagopal}\affiliation{Inspire Institute Inc., Alexandria, Virginia, 22303, USA.}
\affiliation{Harish-Chandra Research Institute, Chhatnag Road, Jhunsi, Allahabad 211 019, India.}
\affiliation{Department of Materials Science \& Engineering, Northwestern University, Evanston, IL 60208, USA.}
\date{\today}

\begin{abstract}
We formulate entropic Leggett-Garg inequalities, which place constraints on  the statistical outcomes of temporal correlations of observables. The information theoretic inequalities are satisfied if {\em macrorealism} holds. We show that the quantum statistics underlying correlations between time-separated spin component  of a quantum rotor mimics  that of  spin correlations  in two spatially separated spin-$s$ particles sharing a state of zero total spin. This brings forth the violation of the entropic Leggett-Garg inequality  by a rotating quantum spin-$s$ system  in similar manner as does the entropic Bell inequality (\prl {\bf 61}, 662 (1988))  by a pair of  spin-$s$ particles forming a composite spin singlet state.    
\end{abstract}
\maketitle

Conflicting foundational features like  non-locality~\cite{Bell}, contextuality~\cite{KS} mark how quantum universe differs from classical one. Non-locality rules out that spatially separated  systems have their own objective properties prior to measurements and they do not get influenced by any local operations by the other parties. Violation of Clauser-Horne-Shimony-Holt (CHSH) - Bell correlation inequality~\cite{CHSH} by entangled states reveals that local realism is untenable in the quantum scenario. On the other hand, quantum contextuality states  that the measurement outcome of an observable depends on the set of compatible observables that are measured alongside it. In this sense, non-locality turns out to be a  reflection of contextuality in spatially separated systems. 

Yet another foundational concept of classical world that is at variance with the quantum description  is {\em macrorealism}~\cite{LG}.  
The notion of {\em macrorealism} rests on the classical world view that (i) physical properties of a macroscopic object
exist independent of the act of observation and (ii) measurements are non-invasive i.e., the measurement of an observable at any  instant of time does not influence its subsequent evolution. Quantum predictions differ at a foundational level from these two contentions. In 1985,  Leggett and Garg (LG)~\cite{LG} designed an inequality (which places bounds on certain linear combinations of temporal correlations of a dynamical observable) to test whether a single macroscopic object exhibits macrorealism or not.  The Leggett-Garg correlation inequality is satisfied by all macrorealistic theories and is violated if quantum law governs. Debates on the emergence of macroscopic classical realm from the corresponding quantum domain continue and it is a  topic of current experimental and theoretical research~\cite{Nature, Nature2, Brukner3, Brukner2, Brukner}.             

Probabilities associated with measurement outcomes in the quantum framework are fundamentally different from those arising in the classical statistical scenario - and this is pivotal in initiating multitude of debates on various contrasting implications in the two worlds~\cite{Fine,Fine2,Pitowski}.  A deeper understanding of these foundational conflicts requires it to be investigated from as many independent ways as possible. The CHSH-Bell (LG) inequalities were originally formulated for dichotomic observables and they constrain certain linear combinations of  correlation functions of spatially (temporally) separated states. However, there have been extensions of correlation Bell inequalities to arbitrary measurement outcomes~\cite{pop}. Information entropy too offers as a natural candidate to capture the puzzling features of quantum probabilities and it offers operational tests demarcating the two domains in an elegant, illustrative fashion~\cite{BC, Fritz2, Kaszlikowski}. The information entropic formulation is applicable to observables with any number of outcomes of measurements. Moreover, while the correlation inequalities define a convex polytope~\cite{Pitowski}, the entropic inequalities form a convex cone~\cite{Yeung}, bringing out their geometrically distinct features.  Entropic tests thus  generalize and strengthen the platform to understand the basic differences between quantum and
classical world view. 

It was noticed quite early by Braunstein and Caves (BC) that interpreting correlations between two spatially separated EPR entangled pair of particles based on Shannon information entropy results in contradiction with local realism~\cite{BC}. They developed information theoretic Bell inequality applicable to any pair of  spatially separated systems and  showed that the inequality is violated by two spatially separated  spin-$s$ particles sharing a state of zero total angular momentum.  More recently, Kurzy{\' n}ski et. al.~\cite{Kaszlikowski} constructed an entropic  inequality  to investigate failure of non-contextuality in a {\em single} quantum three level system and they identified optimal measurements revealing violation of the inequality. Chaves and Fritz~\cite{Fritz} framed a more general entropic framework~\cite{Fritz2} to analyze local realism and  contextuality in quantum as well as post-quantum scenario. Entropic inequalities provide, in general,  a necessary but not sufficient criterion for local realism and non-contextuality~\cite{Fritz, Kaszlikowski}. It is shown that for the $n$-cycle scenario  with dichotomic outcomes, entropic inequalities are also sufficient i.e., the violations of entropic inequalities completely characterize non-local and contextual probabilities in this case~\cite{Chaves, note1}. Application of entropic inequality to test contextuality in four level quantum system has been proposed in Ref.~\cite{pkp}. 

It is highly relevant to address the question "Does the macrorealistic tenet encrypted in the form of classical entropic inequality get defeated in the quantum realm?"  This issue gains increasing importance as  questions on the role of quantum theory in biological molecular  processes  are being addressed in rigorous manner and LG type tests are significant  in recognizing quantum effects in evolutionary biological processes~\cite{MW}. Entropic formulation of macrorealism  generalizes the scope and applicability of such bench-mark investigations. In this paper we formulate entropic LG inequalities to investigate the notion of macrorealism of a single  system.   We show that the entropic inequality is violated by a spin-$s$ quantum rotor (prepared in a completely random state)  in a manner similar to the information theoretic BC inequality for a counter propagating entangled pair of  spin-$s$ particles in a spin-singlet state. To our knowledge, this is the first time that entropic considerations are applied to investigate macrorealism.         

We begin with some basic elements of probabilities and the associated information content  in order to develop the entropic LG inequality in similar spirit as it was formulated by BC~\cite{BC}. Consider a macrorealistic system in which  $Q(t_i)$ is a dynamical observable at time $t_i$. Let the outcomes of measurements of the observable $Q(t_i)$ be denoted by  $q_i$ and the corresponding probabilities $P(q_i)$.   In a macrorealistic theory, the outcomes $q_i$ of   the observables $Q(t_i)$ at all instants of time   pre-exist irrespective of their measurement; this feature is mathematically validated in terms of a joint probability distribution $P(q_1,q_2,\ldots )$ characterizing the statistics of the outcomes; the joint probabilities yield the marginals $P(q_i)$ of individual observations at time $t_i$. Further,  measurement invasiveness implies that the act of observation of $Q(t_i)$ at an earlier time $t_i$ has no influence on its subsequent value at a later time $t_j>t_i$. This demands that the joint probabilities are expressed as a  convex combination of product of probabilities $P(q_i\vert \lambda)$, averaged over a hidden variable probability distribution $\rho(\lambda)$~~\cite{Fine, Brukner}: 
\begin{widetext}
\begin{eqnarray}
\label{ValidProb}
P(q_1,q_2,\ldots , q_n)&=&\sum_{\lambda}\, \rho(\lambda)\, P(q_1\vert \lambda)P(q_2\vert \lambda)\ldots P(q_n\vert \lambda),\\
 \ \ 0\leq \rho(\lambda)\leq 1, \ \sum_\lambda \rho_\lambda=1; && 0\leq P(q_i\vert \lambda)\leq 1,\ \sum_{q_i}\, P(q_i\vert \lambda)=1. \nonumber   
\end{eqnarray}  
\end{widetext}     
Joint Shannon information entropy associated with the measurement statistics of the observable at two different times $t_k, t_{k+l}$ is defined as,  $H(Q_k,Q_{k+l})=-\sum_{q_k,q_{k+l}}\, P(q_k,q_{k+l})\, \log_2\, P(q_k,q_{k+l})$. The conditional information carried by the observable $Q_{k+l}$ at time $t_{k+l}$, given that it had assumed the values $Q_k=q_k$ at an earlier time is given by, 
$H(Q_{k+l}\vert Q_k=q_k)=-\sum_{q_{k+l}}\, P(q_{k+l}\vert q_{k})\, \log_2\, P(q_{k+l}\vert q_{k})$, where  $P(q_{k+l}\vert q_{k})=P(q_{k}, q_{k+l})/P(q_k)$ denotes the conditional probability. The mean conditional information entropy is given by 
\begin{eqnarray}
H(Q_{k+l}\vert Q_{k})&=&\sum_{q_k}\, P(q_k)\,  H(Q_{k+l}\vert Q_k=q_k) \nonumber \\ 
&=& H(Q_k,Q_{k+l})-H(Q_k). 
\end{eqnarray}      
The classical Shannon information entropies obey the inequality~\cite{BC}: 
\begin{equation} 
\label{entin}
H(Q_{k+l}\vert Q_{k})\leq H(Q_{k+l})\leq H(Q_k,Q_{k+l}),
\end{equation}
 left side of which implies that removing a condition never decreases the information -- while right side inequality means that two variables never carry less information than that carried by one of them. Extending (\ref{entin})  to three variables, and also, using the relation $H(Q_k, Q_{k+l})=H(Q_{k+l}\vert Q_{k})+H(Q_{k})$,  we obtain, 
\begin{widetext}
\begin{eqnarray}
\label{h3}
H(Q_k, Q_{k+m})&\leq& H(Q_k, Q_{k+l}, Q_{k+m})=H(Q_{k+m}\vert Q_{k+l}, Q_{k})+ H(Q_{k+l}\vert Q_{k})+H(Q_{k}) \nonumber \\ &&\Longrightarrow 
H(Q_{k+m}\vert Q_k)\leq  H(Q_{k+m}\vert Q_{k+l})+ H(Q_{k+l}\vert Q_{k}). 
\end{eqnarray}
\end{widetext}
Here, the first line follows from the chaining rule for entropies and the derivation is analogous to that given by BC~\cite{BC}.

The entropic inequality (\ref{h3}) is a reflection of the fact that knowing the value of the observable at three different times $t_k<t_{k+l}<t_{k+m}$  -- via its information content --  can never be smaller than the information about it at two time instants.  Moreover, existence of a grand joint probability distribution $P(q_1,q_2,q_3)$ of the variables $Q_1, Q_2, Q_3$, consistent with a given set of marginal probability distributions $P(q_1, q_2)$, $P(q_2, q_3)$, $P(q_1, q_3)$ of pairs of observables, imposes non-trivial conditions on the associated Shannon information entropies. Violation of the inequality points towards lack of a  legitimate grand joint probability distribution for all the measured observables, such that the family of probability distributions associated with measurement outcomes of  pairs of observables belong to it as marginals~\cite{Fritz2, note}.

The same reasoning, which lead to a three term entropic inequality (\ref{h3}), could be extended to construct  an entropic inequality for  $n$ consecutive measurements  $Q_1, Q_2, \ldots, Q_n$ at time instants  $t_1<t_2<\ldots <t_n$: 
\begin{widetext}
\begin{equation}
\label{hn}
H(Q_n\vert Q_1)\leq H(Q_n\vert Q_{n-1})+H(Q_{n-1}\vert Q_{n-2})+\ldots +H(Q_{2}\vert Q_{1}).
\end{equation} 
\end{widetext}
The macrorealistic information underlying the  statistical outcomes of the observable at $n$ different times must be consistent with the information associated with pairwise non-invasive  measurements as given in (\ref{hn}).

Note that for even values of $n$, there is a one-to-one correspondence between the entropic inequality (\ref{hn}) of a single system  and the information theoretic BC inequality~\cite{BC} for two spatially separated parties (Alice and Bob). More specifically, let us consider $n=4$ in (\ref{hn}) and  associate temporal observable $Q_i$ with Alice's (Bob's) observables $A',\ A$  ($B',\ B$)  as   $Q_1\leftrightarrow B$, $Q_2\leftrightarrow A'$, $Q_3\leftrightarrow B'$, $Q_4\leftrightarrow A$ to obtain the BC inequality~\cite{BC} for a set of  four correlations: $H(A\vert B)\leq H(A\vert B')+ H(B'\vert A')+H(A'\vert B)$, which is satisfied by any {\em local realistic}  model of spatially separated pairs. It may be identified that Eq.~(\ref{ValidProb}) is essentially analogous  to local hidden variable model (Bell scenario for spatially separated systems) as well as  non-contextual model, while the interpretation here is towards  macrorealism. Moreover, we emphasize that the logical reasoning in formulating the entropic LG inequalities (\ref{hn}) is synonymous to that of BC~\cite{BC}, which indeed offers a unified approach to address non-locality, contextuality and also non-macrorealism.              

We proceed to show that LG entropic inequality is  violated by a quantum spin-$s$ system. Consider a quantum rotor prepared initially in a maximally mixed state
\begin{equation}
\label{rhoin}
\rho=\frac{1}{2s+1}\sum_{m=-s}^{s}\ \vert s,m\rangle\langle s,m\vert =\frac{I}{2s+1}  
\end{equation}         
where $\vert s, m\rangle$ are the simultaneous eigenstates of the squared spin operator $S^2=S_x^2+S_y^2+S_z^2$ and the $z$-component of spin $S_z$ (with respective eigenvalues $ s(s+1)\, \hbar^2$ and $m\hbar$); $I$ denotes the $(2s+1)\times (2s+1)$ identity matrix. We consider the Hamiltonian 
\begin{equation}
H= \omega\, S_y, 
\end{equation}
resulting in  the unitary  evolution $U(t)=e^{-i\omega t\, S_y/\hbar}$ of the system (which corresponds to a rotation about the $y$-axis by an angle $\omega\, t$).  We choose  $z$-component of spin $Q(t)=S_z(t)=U^\dag(t)\, S_z\, U(t)$ as the dynamical observable for our investigation of macrorealism. Let us suppose that the observable $Q_k=S_z(t_k)$ takes the value $m_k$ at time $t_k$. Correspondingly, at a later instant of time $t_{k+l}$ if the spin component $S_z(t_{k+l})$ assumes the value $m_{k+l}$, the quantum mechanical joint probability is given by~\cite{Brukner} 
\begin{equation}
P(m_k, m_{k+l})=P_{m_k}(t_k)\, P(m_{k+l}, t_{k+l}\vert m_{k}, t_{k}).
\end{equation}  
Here,  $P_{m_k}(t_k)={\rm Tr}[\rho\, \Pi_{m_k}(t_k)]$ is the probability of obtaining the outcome $m_k$ at time $t_k$,  
$P(m_{k+l}, t_{k+l}\vert m_{k}, t_{k})={\rm Tr}[\Pi_{m_k}(t_k)\rho\Pi_{m_k}(t_k)\, \Pi_{m_{k+l}}(t_{k+l})]/P_{m_k}(t_k)$ denotes the conditional probability of obtaining the outcome $m_{k+l}$ for the spin component $S_z$ at time $t_{k+l}$, given that it had taken the value $m_k$ at an earlier time $t_k$; 
$\Pi_{m}(t)=U^\dag(t)\, \vert s, m\rangle\langle s, m\vert\, U(t)$ is the projection operator measuring the outcome $m$ for the spin component at time $t$.        For the maximally mixed initial state (\ref{rhoin}), we obtain the quantum mechanical  joint probabilities as, 
\begin{eqnarray}
\label{cp}
P(m_k, m_{k+l})&=&\frac{1}{2s+1}\, {\rm Tr}[\Pi_{m_k}(t_k)\, \Pi_{m_{k+l}}(t_{k+l})]\nonumber \\
&=&\frac{1}{2s+1}\,  \vert \langle\, s, m_{k+l}\vert e^{-i\omega (t_{k+l}-t_k)\, S_y/\hbar}\,\vert s, m_{k}\rangle\vert^2\nonumber \\ 
&=& \frac{1}{2s+1}\,  \vert\, d^{s}_{m_{k+l}\, m_k}(\theta_{kl})\vert^2  
\end{eqnarray}      
where $d^{s}_{m' m}(\theta_{kl})=\langle s, m'\vert e^{-i\theta_{kl}\, S_y/\hbar}\vert s, m\rangle$ are the matrix elements of the $2s+1$ dimensional irreducible representation of rotation~\cite{Rose}  about $y$-axis by an angle $\theta_{kl}=\omega (t_{k+l}-t_k)$. The marginal probability of the outcome $m_k$ for the observable $Q_k$ is readily obtained by making use of the unitarity property of $d$ matrices:
$P(m_k)=\sum_{m_{k+l}}\, P(m_k, m_{k+l})=\frac{1}{2s+1}\,  \sum_{m_{k+l}}\vert\, d^{s}_{m_{k+l}\, m_k}(\theta_{kl})\vert^2=\frac{1}{2s+1}.$    

Clearly, the temporal correlation probability (\ref{cp}) of quantum rotor is similar to the quantum mechanical pair probability~\cite{BC} 
\begin{eqnarray}
P(m_a, m_b)&=&\left[_{\hat{a}}\langle s,m_a\vert \otimes _{\hat{b}}\langle s,m_b \vert \right]\ \vert \Psi_{AB}\rangle \nonumber \\
&=& \frac{1}{2s+1}\, \vert d^s_{m_a,-m_b}(\theta_{ab})\vert^2
\end{eqnarray}
that Alice's measurement of spin component $\vec{S}\cdot \hat{a}$  yields the value $m_a$ and Bob's measurement of  $\vec{S}\cdot \hat{b}$  results in the outcome  $m_b$ in a spin singlet state $\vert \Psi_{AB}\rangle=\frac{1}{\sqrt{2s+1}}\, \sum_{m=-s}^{s}\, (-1)^{s-m}\ \vert s,m\rangle \otimes \vert s,-m\rangle$ of a spatially separated pair of spin-$s$ particles. (Here $\theta_{ab}$ is the angle between the unit vectors $\hat{a}$ and $\hat{b}$). In other words,  quantum statistics of temporal correlations in a single spin-$s$ rotor  mimics that of  spatial correlations in an entangled counter propagating pair  of  spin-s particles.     

Let us consider measurements at equidistant time intervals $\Delta t=t_{k+1}-t_k, \ k=1,2,\ldots n$ and denote $\theta=(n-1) \omega\, \Delta t$. The quantum mechanical information entropy depends only on the time separation, specified by the angle $\theta$ and is given by,  
\begin{widetext}
\begin{eqnarray}
H(Q_k\vert Q_{k+1})\equiv H[\theta/(n-1)]&=&-\frac{1}{2s+1}\, \sum_{m_k, m_{k+1}}\, \vert d^s_{m_{k+1},m_{k}}[\theta/(n-1)]\vert^2 \log_2 \vert d^s_{m_{k+1},m_{k}}[\theta/(n-1)]\vert^2.
\end{eqnarray}    
The $n$-term entropic inequality (\ref{hn}) for observations at equidistant time steps assumes the form, 
\begin{eqnarray}
\label{ed}
(n-1)\, H[\theta/(n-1)]-H(\theta)&=& -\frac{1}{2s+1}\sum_{m_k, m_{k+1}}\left[\,(n-1) \, \vert d^s_{m_{k+1},m_{k}}[\theta/(n-1)]\vert^2\, \log_2 \vert d^s_{m_{k+1},m_{k}}[\theta/(n-1)]\vert^2\right.   \nonumber \\
&& \left. - \vert d^s_{m_{k+1},m_{k}}(\theta)\vert^2\, \log_2 \vert d^s_{m_{k+1},m_{k}}(\theta)\vert^2\right]\geq 0
\end{eqnarray} 
 \end{widetext}
We introduce  information deficit, measured in units of $\log_2 (2s+1)$ bits, as 
\begin{equation}
\label{dn}
{\cal D}_n(\theta)= \frac{(n-1)\, H[\theta/(n-1)]-H(\theta)}{\log_2 (2s+1)}
\end{equation} 
so that the  violation of the LG entropic inequality (\ref{ed}) is implied by negative values of ${\cal D}_n(\theta)$. 
The units $\log_2 (2s+1)$ for the quantity  ${\cal D}_n(\theta)$ imply that the base of the logarithm for evaluating the entropies of a spin $s$ system is chosen appropriately to be $(2s+1)$. For a spin-1/2 rotor, it is in bits.       

In Fig.~1, we have plotted information deficit ${\cal D}_n(\theta)$ for $n=3$ (Fig.~1a) and $n=6$ ~(Fig.~1b) as a function of  $\theta=(n-1)\, \omega\, \Delta t$  for spin values $s=1/2, 1, 3/2$ and 2. The results illustrate that the information deficit   assumes negative values, though the  range of violation  (i.e., the value of the angle $\theta$ for which the violation occurs) and also the strength (maximum negative value of $D_n(\theta)$) of the entropic violation   reduces~\cite{noteusha} with the increase of spin $s$.  This implies the emergence of macrorealism for the dynamical evolution of a quantum rotor in the limit of large spin $s$. It may be noted that Kofler and Brukner~\cite{Brukner2} had  shown,  violation of the correlation LG inequality -- corresponding to the measurement outcomes of a dichotomic parity observable in the example of a quantum rotor -- persists  even for large values of spin if the eigenvalues of spin can be experimentally resolved by sharp quantum measurements. However, under the restriction of coarse-grained measurements classical realm emerges in the large spin limit.

\begin{figure}[h]
\includegraphics*[width=2.4in,keepaspectratio]{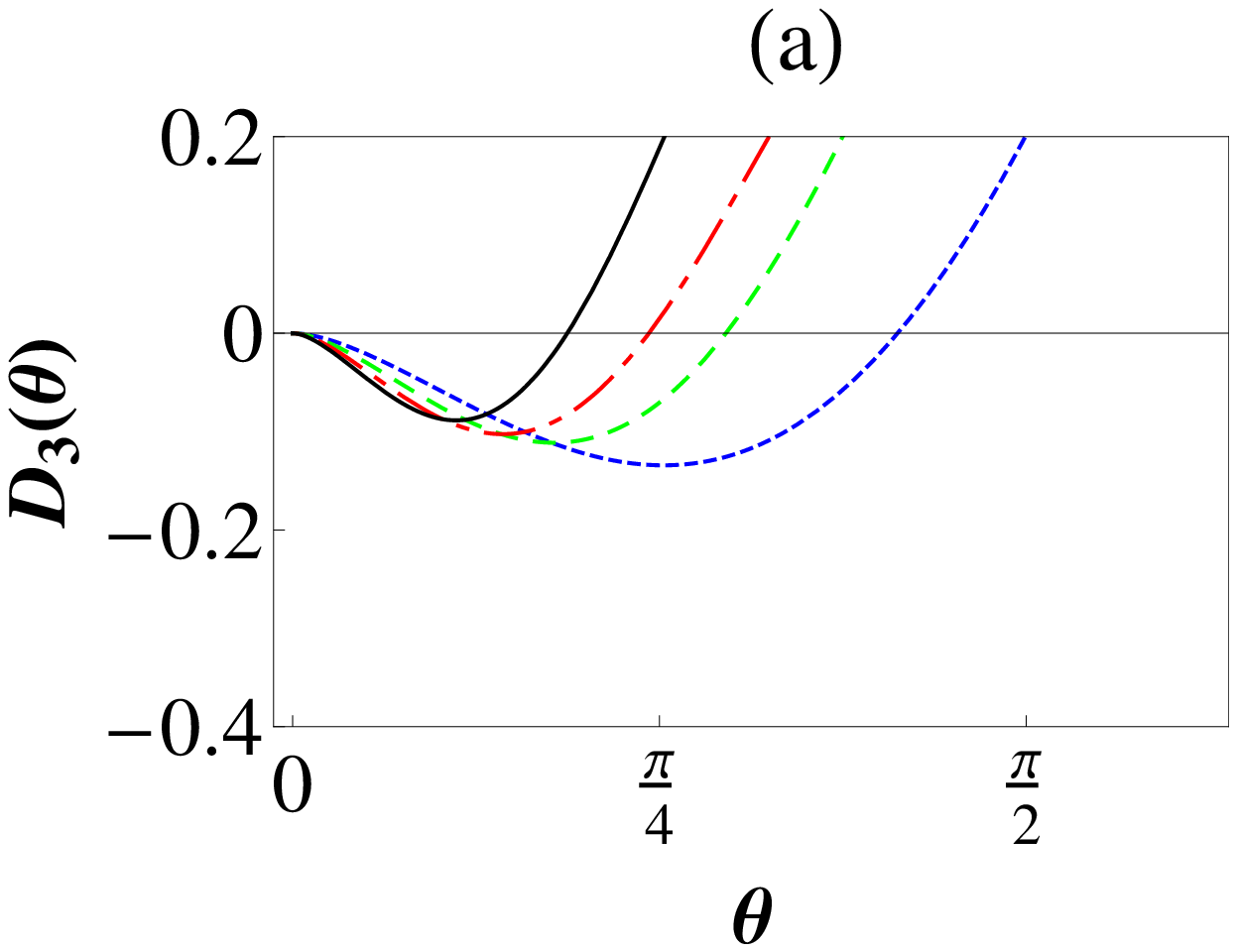}
\includegraphics*[width=2.4in,keepaspectratio]{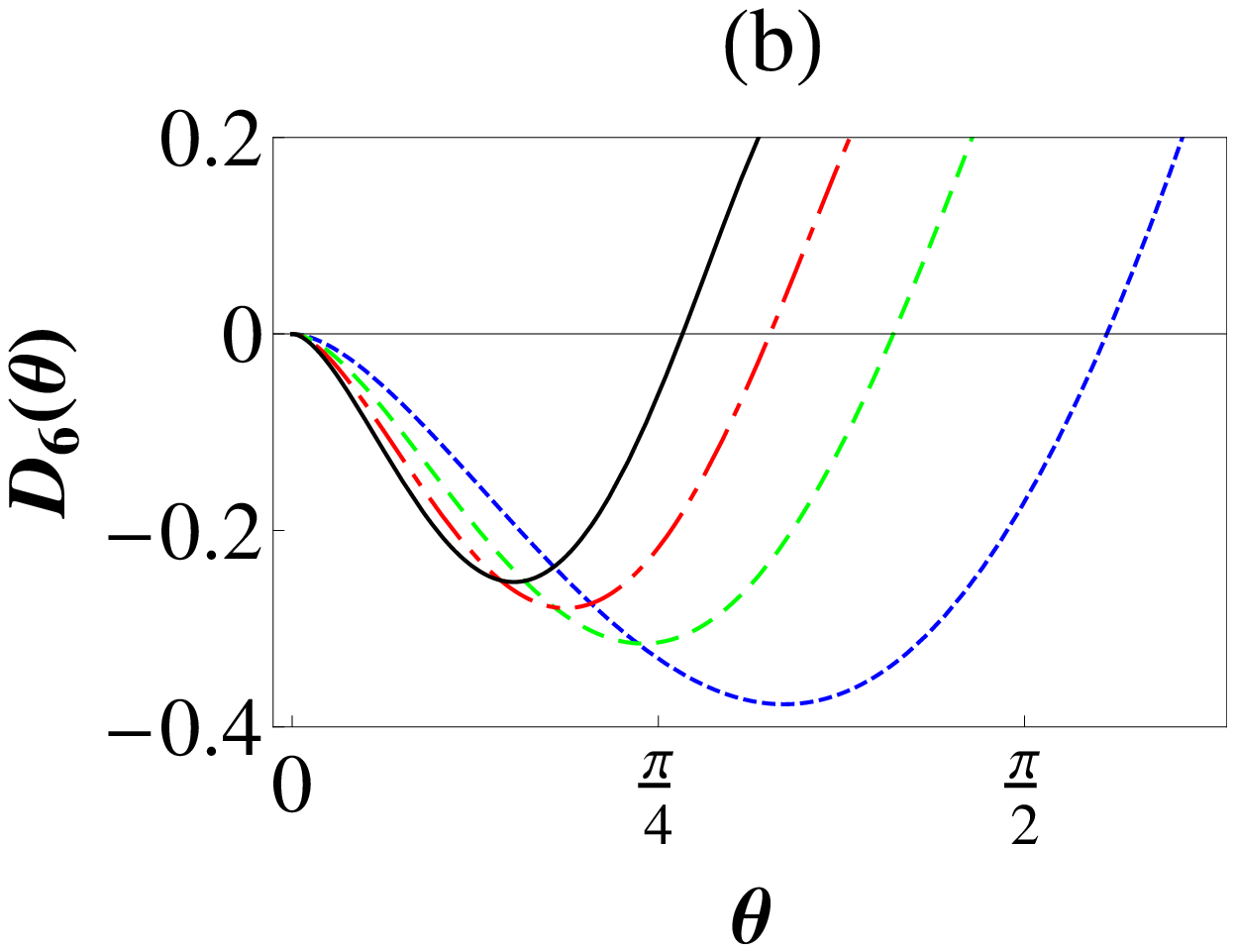}
\caption{(Color online) LG Information deficit ${\cal D}_n(\theta)$  of (\ref{dn})) -- in units of $\log_2(2s+1)$ bits --  corresponding to the measurement of the spin component $S_z(t)$ of a quantum rotor, at equidistant time steps $\Delta t=\frac{\theta}{(n-1)\, \omega }$,   number of observations being (a) $n=3$ and (b) $n=6$ during the total time interval specified by the angle $\theta=(n-1)\, \omega\, \Delta t$.  Conflict with macrorealism is recorded  by the negative value of ${\cal D}_n(\theta)$. Maximum negative value and also the range i.e., the value of $\theta$
 over which the information deficit is negative,   grows with the increase in the number $n$ of observations. However, for a given $n$, both the strength and the range of violation  reduce with the increase of spin value (spin-1/2: dotted; spin-1: dashed; spin-3/2: dot-dashed; spin-2: solid curve). The strength of violation may be related to how much inconsistent  Shannon information entropies could be -- when the associated probabilities of outcomes of pairs of dynamical observables have their origin in {\rm noisy} quantum  measurements  -- compared to those arising within a classical macrorealistic premise. All quantities are dimensionless.}
\end{figure}
      
Macrorealism requires that a consistently larger information content $H[\theta/(n-1)]$ has to be carried by the system, when number  of observations $n$ is increased and  small steps of time interval are employed; however, quantum situation does not comply with this constraint. More specifically, in the classical premise, knowing the observable at almost all time instants provides more information content, whereas, quantum realm results in less information with large number of observations. To see this explicitly, consider the limit of  $n\rightarrow \infty$ and infinitesimal time steps  $\omega\, \Delta t\rightarrow 0$. Quantum statistics leads to vanishingly small information  i.e., $H(\frac{\theta}{n-1})\rightarrow 0$ --  a signature of quantum Zeno effect. In this limit,  the information deficit (see (\ref{dn}))   $D_n(\theta)\rightarrow \frac{-H(\theta)}{\log_2 (2s+1)}$ is  negative -- thus violating the entropic LG inequality. The entropic test clearly brings forth the severity of  macrorealistic demands towards {\em knowing} the observable in a  non-invasive manner under such miniscule time scale observations. 

In conclusion, we have formulated entropic LG inequality, which places bounds on the amount of information associated with non-invasive measurement of a macroscopic observable.  The entropic formulation can be applied to any observables -- not necessarily dichotomic ones -- and it puts to test macrorealism i.e.,  a combined demand of the  pre-existence of definite values of the measurement outcomes of a given  dynamical observable at different instants of time --  together with the assumption that act of observation at an earlier instant does not influence the subsequent evolution. The information entropic approach provides a unified approach to test local realism, non-contextuality and macrorealism.  

The classical notion of macrorealism demands that  statistical outcomes of measurement of an observable at consecutive time intervals originate from a valid grand joint probability,  presumably of the form (\ref{ValidProb}). Non-existence of a legitimate joint probability, such that the family of probability distributions associated with the measurement outcomes of every pair of observables belong to it as marginals, reflects through the violation of the entropic test.  The violation also brings forth the fact that more information is associated with the knowledge of the observable at more instants of time in the classical macrorealistic realm -- however, more number of observations correspond to less information in the quantum case.  

In order to demonstrate violation of the entropic inequality, we considered the dynamical evolution of a quantum spin system prepared initially in a maximally mixed state. We have demonstrated that the entropic violation in a  quantum rotor system is similar to that of a spatially separated pair of spin-$s$ particles sharing a state of total spin zero~\cite{BC}. Further, we have illustrated that the information content of a rotor grows with the increase of spin $s$ such that it is consistent with the requirements of macrorealism.

We thank T. S. Mahesh and Hemant Katiyar for sharing their experimental results through private communication and for several insightful discussions. 
   
\noindent {\em Note added}: After submission of this paper, Hemant et. al.~\cite{Hemant} have reported experimental violation of entropic LG inequalities in an ensemble of spin 1/2 nuclei using nuclear magnetic resonance (NMR) techniques, by recording negative values of information deficit ${\cal D}_3$ --  in striking agreement with our theoretical prediction. Further, they have demonstrated that the experimentally extracted three time joint probabilities do not contain all the pairwise probabilities as marginals  -- which reflects the failure of the entropic test (see \cite{note}).


\begin{thebibliography}{0}
\bibitem{note}  Consider sequential measurement of the observables $Q_1$, $Q_2$ and $Q_3$ resulting in  the measurement outcomes  $q_1, q_2, q_3$ in a quantum system prepared in the state $\rho$. With measurements of only  $Q_1, Q_2$, the probabilities $P(q_1,q_2)=P(q_1)\, P(q_2\vert q_1)$ with $P(q_2\vert q_1)=Tr[\Pi_{q_2}\, \rho_{q_1}]$, where $\rho_{q_1}=[\Pi_{q_1} \rho\, \Pi_{q_1}]/P(q_1);\ P(q_1)={\rm Tr}[\rho\, \Pi_{q_1}]$.  A consequent measurement of $Q_3$ yields the probabilities $P(q_1,q_2,q_3)=P(q_1, q_2)\, P(q_3\vert q_1, q_2); P(q_3\vert q_1, q_2)={\rm Tr}[\Pi_{q_3}\, \rho_{q_1\,q_2}]$ and
$\rho_{q_1\,q_2}=\Pi_{q_2}\, \rho_{q_1}\, \Pi_{q_2}/P(q_1,q_2)$. After simplification, the three variable joint probabilities reduce to the form, 
$P(q_1,q_2,q_3)=P(q_1,q_2)\, P(q_2,q_3)/P(q_2)$.  It is easy to see that the three variable probabilities $P(q_1,q_2,q_3)$ associated with the sequential measurement of the observables $Q_1,\ Q_2,$ and  $Q_3$ are not consistent with the pairwise probabilities $P(q_1,q_3)$. 
Absence of a legitimate grand joint probability distribution, consistent with all pairwise probabilities reflects itself in the violation of entropic inequality~\cite{Fritz2}.
\end{thebibliography}
\end{document}